\documentclass[letterpaper]{article} 
\usepackage{aaai2026}  
\usepackage{times}  
\usepackage{helvet}  
\usepackage{courier}  
\usepackage[hyphens]{url}  
\usepackage{graphicx} 
\usepackage{natbib}  
\usepackage{caption} 
\usepackage{algorithm}
\usepackage{algorithmic}
\usepackage{soul}
\usepackage{booktabs}
\usepackage[T1]{fontenc} 

\usepackage{dblfloatfix}
\usepackage{xcolor} 

\usepackage{newfloat}
\usepackage{listings}
\DeclareCaptionStyle{ruled}{labelfont=normalfont,labelsep=colon,strut=off} 
\floatstyle{ruled}
\newfloat{listing}{tb}{lst}{}
\floatname{listing}{Listing}
\title{Between Commons and Property: Individual Image-Generation Model Creators' Perceived Accountability for Derivative AI Model on Open Model Marketplaces}
\author{
    Eun Jeong Kang\textsuperscript{\rm 1}, Fengyang Lin\textsuperscript{\rm 1}, Angel Hsing-Chi Hwang\textsuperscript{\rm 2}
}
\affiliations{
    \textsuperscript{\rm 1}Cornell University,\\
    \textsuperscript{\rm 2}University of Southern California\\

    eunkang@infosci.cornell.edu, fl354@cornell.edu,
    angel.hwang@usc.edu
}

\usepackage{bibentry}

\begin{document}

\maketitle

\begin{abstract}
  Within the open model ecosystem, AI model marketplaces enable non-AI experts to create and distribute derivative AI models. These platforms establish RAI approaches that creators are expected to follow to prevent negative downstream effects from model deployment, yet such governance approaches often rely on creators' intrinsic motivation and integrity. Little is known about how individual creators conceptualize their accountability for derivative models and how these conceptualizations shape their RAI practices throughout the development cycle.

Through semi-structured interviews with 19 creators of derivative image-generation models on open model marketplaces, we identify two distinct accountability logics underlying creators' development practices: commons-based peer production and personal creative property. Guided by these logics, creators engage with platform RAI approaches in ways that diverge from their intended purposes: they strategically repurpose attribution tools for model visibility rather than provenance transparency, displace platform authority in favor of community-based governance, restrict model access to manage downstream accountability costs, and treat RAI-related work as undervalued labor to be minimized.
We discuss how self-regulatory use of RAI tools, shaped by individuals' perceptions, produces shortcomings, and how open model marketplaces might mitigate these limitations by structurally supporting creators' engagement with RAI practices.

\end{abstract}


\section{Introduction}
Open AI model marketplaces \footnote{In this paper, we use ``open AI'' as an umbrella term for open-source and open-weight models, following Widder et al. \cite{widder2023open}.}, such as Hugging Face and Civit AI, act as critical intermediate ML/AI model service providers in the open model ecosystem, contributing to innovating AI development pipelines~\cite{widder2023open, raffel2023building, gorwa2024moderating}. 
These platforms enable users to easily access user-uploaded models as well as create derivatives from those models~\cite{choksi2025brief}. 
Individuals -- including hobbyists and independent developers -- can build customized derivative models without training from scratch or a lot of resources, through fine-tuning open-source AI models, custom dataset training, or other modification methods~\cite{anisuzzaman2025fine, sola2026}.

As the number of models hosted on the platforms grows, concerns regarding downstream effects of user-created custom models have intensified~\cite{schneider2025investigating, wei2024exploring}. 
Pre-trained models can be fine-tuned on copyrighted content or pornographic imagery to serve individuals' personal interests. 
In 2025, over 5,000 models generating non-consensual sexual imagery of real individuals were reportedly migrated to Hugging Face from another major marketplace~\cite{maiberg2025huggingface}.
Once these models are downloaded by downstream users, creators can no longer trace how users interact with their models or how those models spread~\cite{widder2022limits}.
These concerns are particularly pronounced for image generation models, where the aesthetic visual content amplifies both public attention and potential harms~\cite{marchal2024generative, lima2025public}. 
Because such harms often remain latent until downstream deployment, platforms increasingly need governance approaches that extend beyond moderation of within-platform trust and safety issues.

Platforms have begun to promote responsible open model development guidelines and Responsible AI (RAI) tools for mitigating harms that creators may embed in their models~\cite{gorwa2024moderating, berman2024scoping, gebru2021datasheets}. Beyond centralized moderation, these approaches are intended to support creators in voluntarily upholding the safety and transparency of their models~\cite{mitchell2019model, he2025contributions}.

However, these existing self-regulatory approaches yield limited effect, particularly when creators lack intrinsic motivation or relevant expertise. Creators apply ad-hoc, self-constructed ethical norms drawn from personal experience when developing models without a concrete understanding of the actors involved in the AI pipeline~\cite{widder2022limits, widder2023dislocated}. 
Without a standardized legal framework enforcing RAI practices~\cite{hacker2026regulation}, creators tend to enact RAI approaches according to their own sense of accountability for the models they produce. For instance, model card completion relies on creators' integrity and accountability~\cite{mitchell2019model}; documents may be left blank or filled minimally when creators anticipate that detailed documentation could enable malicious actors to exploit their models. 
However, we have limited understanding of how individual creators conceptualize their accountability for their models -- open derivative models customized through model modification methods -- and how this conceptualization, together with their broader values and community relationships, shapes their RAI practices throughout the development cycle.

The goal of this work is therefore twofold: (1) to examine how existing platform governance approaches become misaligned with creators' experiences, and (2) to envision new approaches to platform governance grounded in individual creators' practices and their experiences on these platforms, in ways that better support responsible AI.
Through semi-structured interviews with 19 individual model creators who produce and distribute image-generation derivative models on open model marketplaces, we examine their current model creation and distribution practices and how platform governance approaches influence such processes.
To generate concrete feedback and envision alternative governance approaches, we gather input from model creators on how they respond to existing Responsible AI (RAI) approaches (such as responsible model development guidelines and attribution tools) used in other AI-infused creative contexts.
Together, our interviews address two key research questions:

\begin{enumerate}
\item \textbf{RQ1.} How do individual model creators' perspectives on image-based derivative models shape their sense of accountability in model development and sharing?
\item \textbf{RQ2.} How do individual model creators engage with RAI approaches on open model marketplaces, and in what ways do their practices diverge from the intended purposes of these approaches?
\end{enumerate}

Our findings reveal two distinct logics underlying how model creators conceptualize accountability for their derivative models, each producing a different orientation toward responsible model development. The first logic draws on commons-based peer production, where creators view their work as a byproduct of community engagement and shared resources~\cite{benkler2006commons}. The second draws on personal creative property, where creators treat their derivatives as personal proprietary artifacts~\cite{birch2020assetization}. These logics interact, jointly guiding creators' derivative model production cycle.

Within these logics, RAI approaches are practiced based on creators' understanding of how responsible practices fit their circumstances on platforms, within their communities, and in relation to themselves.
When engaging with RAI approaches provided by platforms, creators' practices diverge from the intended purposes of these tools in several ways: they strategically repurpose attribution tools for model visibility rather than provenance transparency, rely on community-based governance that displaces platform authority, restrict model access to manage downstream accountability costs, and treat RAI-related work as undervalued labor to be minimized.
These practices, while locally rational for creators, can distort the transparency of model dependencies and complicate attribution across the open model ecosystem.

Together, our findings show that platform governance approaches that rely on individuals' decisions self-regulating their workflow are insufficient, given individual creators' divergent motivations, situated perspectives, and constantly changing perceived accountability. We argue that addressing these risks requires recognizing individual creators as a structural locus of risk, and suggest mechanisms by which platforms can support individuals in performing meaningful RAI practices.

\section{Background and Related Work}

\subsection{The Role of Open Model Marketplaces in the Open Model Ecosystem}
Aligned with the open-source culture -- in which software is freely distributed, technologically neutral, and granted free sub-licensing rights~\cite[p. 577]{aksulu2010comprehensive} -- that has historically fostered innovation and democratized software development~\cite{meneely2009secure, googlea}, openness in AI has been strongly endorsed by scholars and industry actors as a means to stimulate innovation and experimentation while strengthening accountability \cite{tan2025if, paris2025opening, langenkamp2022open, kilamo2012proprietary}.
This openness also facilitates community engagement through peer-production -- such as collaboratively auditing code and involving stakeholders in value-centered development, which has been viewed as a pathway toward more affordable, transparent, and accountable AI systems~\cite{eiras2024position, widder2024open}.

In line with these goals, open model ecosystems have emerged with multiple actors operating across the AI supply chain. Major AI corporations and organizations (e.g., Meta, Stability AI~\footnote{https://stability.ai/}) publicly released pre-trained models under open licenses for non-commercial purposes~\cite{industry, eiras2024position, liesenfeld2024rethinking}.
Building on these open models, derivatives -- models customized through post-training modification methods such as fine-tuning -- can be created and hosted on dedicated platforms, which we refer to as \textit{``open model marketplaces''} (e.g., Hugging Face, CivitAI)~\cite{choksi2025brief}. Within this ecosystem, our research focuses on open model marketplaces and the users who create and distribute derivative models on these platforms.

\subsection{Governing Open Model Marketplaces for Responsible AI}

Open model marketplaces are known for affording users freedom to experiment with hosted models, but researchers have raised concerns that this openness can also facilitate the distribution of models that produce harmful outputs or operate in unresolved legal areas~\cite{harris2023regulate, wei2024exploring, maiberg_huggingface_2025}. 
First, models fine-tuned on non-consensual intimate images (NCII) that violate publicity rights can be widely distributed via the platforms~\cite{wei2024exploring, darka}. Researchers found that Text-to-Image models hosted on these marketplaces can present representational harms and biases when downstream users generate images, calling for strategies to mitigate downstream harms~\cite{schneider2025investigating}.
Additionally, these models carry varied licenses that are not always complied with across derivative chains~\cite{duan2025position, duan2024they}.

Under legal pressure to ensure internet trust and safety, platforms have attempted to regulate models that generate child sexual abuse material (CSAM)-related content or that may amplify harms when used by downstream users~\cite{ykilcher, gorwa2024moderating}.
For instance, CivitAI removed all user-uploaded models designed to generate CSAM or NCII of real individuals~\cite{maiberg2025payment, maiberg2025huggingface}.
However, platforms encounter significant governance challenges in moderating these models, in part because accountability is difficult to allocate across chains of derivative models~\cite{laufer2025anatomy}. The dual-use nature of these models further complicates moderation, as platforms must distinguish legitimate creative uses from unintended harmful applications~\cite{gorwa2024moderating, franccois2025different}. Centralized moderation at the platform level is thus likely to incur substantial operational costs while still overlooking latent risks that only manifest in downstream contexts.

As a complementary approach, platforms promote user-driven, Responsible AI (RAI) practices through which users self-regulate their model development. These approaches typically rely on individuals' own sense of moral responsibility, assuming that actors will act in good faith. One mechanism is the \textit{responsible model development guideline,} which describes what data should be filtered during fine-tuning and which kinds of outputs creators should be cautious about when deploying their models. Another approach is \textit{attribution disclosure,} through which creators use templates or tools to document their evaluation results, data sources, and licenses~\cite{mitchell2019model, gebru2021datasheets, pepe2024hugging}. These provisions are intended to enhance the transparency, reproducibility, and accountability of derivative models, helping downstream users understand how to apply them. 

However, prior work has demonstrated that ``Responsible AI (RAI)'' is not universally defined or practiced. Depending on the purpose and activities of AI development, people hold different priorities and norms about what to prioritize as their responsibility~\cite{jakesch2022different, widder2022limits, widder2023dislocated, sadek2025challenges}. Moreover, prior research highlights that RAI approaches (e.g., Model Cards, Data Cards) may not be fully implemented by AI practitioners when other concerns, such as reputational and political risks, take precedence~\cite{kawakami2024responsible}. Indeed, model documentation practices on one of the major model marketplaces (Hugging Face) remain insufficient, and cooperative commitments to transparency and reproducibility across open models are weakly upheld~\cite{choksi2025brief, pepe2024hugging}. 

Bovens argues that accountability can be assigned through specific mechanisms or social relations that involve an obligation to explain conduct~\cite{bovens2010two}. That is, how actors are rewarded for their creations, how they interact with other actors (e.g., community members), and to whom they are held accountable may affect how they use RAI approaches. The users of open model marketplaces include not only AI practitioners affiliated with companies or institutions, but also individuals who create AI models recreationally and independent freelancers. Widder et al. show that these creators enact ethical norms shaped by personal beliefs and cultural and labor-related experiences, often relying on informal community guidelines in open-source AI projects~\cite{widder2022limits}. Personal motivations such as reputational concerns further drive how they scrutinize their work~\cite{bartsch2024origin}. Tang et al. further show that developers experience epistemic, methodological, and ecosystem-level uncertainties when documenting their open-source models, pointing to the importance of infrastructural support across the AI supply chain~\cite{tang2025navigating}. Yet little is known about how platform governance approaches---which require creators to self-regulate their workflows through RAI practices---function for individuals accessing, using, or distributing derivative models on these platforms. 

We therefore examine how these creators conceptualize their accountability from their model development practices, and then how they engage with RAI approaches in their workflows.

\begin{table*}
\resizebox{0.98\textwidth}{!}{%
\renewcommand{\arraystretch}{1.2}
    \begin{tabular}{p{0.08\linewidth}p{0.1\linewidth}p{0.25\linewidth}p{0.22\linewidth}p{0.35\linewidth}}
        \toprule
        \textbf{PID} & \textbf{Gender} & \textbf{\shortstack[l]{Professional Context}} & \textbf{\shortstack[l]{Concept of \\ Image-generation models}} & \textbf{\shortstack[l]{Platforms Where \\ Participants Deploy Derivative Models}} \\
        \hline
        P1 & Male & Hair designer & Afro hair style texture & Github \\
        \hline
        P2 & Female & Creative technologist & Graphic visual & Hugging Face, CivitAI\\
        \hline
        P3 & Male & UI/UX designer & Japanese anime character & Discord (Private) \\
        \hline
        P4 & Male & UI/UX designer & Sports player meme & CivitAI \\
        \hline
        P5 & Male & Graphic designer & Afro-artistic image & CivitAI \\
        \hline
        P6 & Male & Independent model creator & Realistic portrait & CivitAI \\
        \hline
        P7 & Male & Graduate (CS) & Japanese anime character & CivitAI\\
        \hline
        P8 & Male & Graduate (CS) & Anime character fashion & Discord (Private) \\
        \hline
        P9 & Male & Engineer & Superhero anime character & Discord (Private)  \\
        \hline
        P10 & Male & Undergraduate (CS) & Asian culture & Hugging Face, CivitAI \\
        \hline
        P11 & Female & Graduate (CS) & Graphic visual & Discord (Private) \\
        \hline
        P12 & Male & Independent model creator & Photorealistic character & Hugging Face \\
        \hline
        P13 & Male & UI/UX designer & Photorealistic portrait & Hugging Face, CivitAI \\
        \hline
        P14 & Male & Engineer & 80s fashion style & CivitAI \\
        \hline
        P15 & Male & Graphic designer & Animal character  & CivitAI, Github \\
        \hline
        P16 & Female & Physiotherapist & Education materials & Discord (Private) \\
        \hline
        P17 & Male & UI/UX designer & Japanese anime character  & Discord (Private) \\
        \hline
        P18 & Male & Computer scientist & {Japanese anime character,\newline photorealistic portraits}   & Hugging Face, CivitAI \\
        \hline
        P19 & Prefer not to say & Engineer & Japanese anime character &  Hugging Face , CivitAI \\
        \bottomrule
    \end{tabular}%
}
\vspace{0.5em}
\caption{Demographic Information of Participants. Creative concept of model is reported based on the concept of image output generated from a model.}
\label{tab:participants}
\end{table*}

\section{Methodology}
We conducted semi-structured interviews with 19 individual creators who create and distribute derivative image-generation models to understand how they conceptualize their accountability over their workflows and how they engage with RAI approaches in their practices. The study was approved by our University Institutional Review Board (IRB), and participants received a \$35 gift card for their 75–90 minute sessions.

\subsection{Participants}
We define individual model creators as users of online platforms who create and distribute open models independently. We focus on creators of image-generation models, since this type of model has attracted significant public attention as a form of entertainment, alongside concerns about its use to produce sexually abusive content~\cite{marchal2024generative, lima2025public}. 
Criteria also included being over 18 years of age and residing in the United States.

Participants were recruited via social media channels (LinkedIn, Reddit, Discord). Due to the limited population, we also reached out on model marketplaces (Hugging Face, CivitAI, and Tensor.Art) using their public nomination lists and snowball sampling~\cite{parker2019snowball, choksi2025brief}.

\subsubsection{Demographic Information}
Nineteen individual model creators participated in the study (see Table~\ref{tab:participants}). Although we did not restrict recruitment to creators of derivative models specifically, all participants developed \textit{derivative models}, open models customized from pre-trained models through modification methods such as fine-tuning. Most participants distributed their models on Hugging Face and CivitAI, while six participants used private Discord channels for sharing models.

Seventeen participants were hobbyists, while two (P6, P12) identified themselves as independent freelance engineers, who operated independently and received commissions for building custom models. P2 has developed only image-to-image models, while P18 has created both image-to-image and text-to-image models. All others have only produced text-to-image models. Participants developed derivative models by re-using pre-trained model weights using either node-based interfaces like Comfy-UI or code-based environments like Google Colab. All participants' models are derivatives of Stable Diffusion or Flux open weight models. 

We also acknowledge the gender imbalance among participants. 
We suspect that the predominance of anime character models on model marketplaces~\cite{song2024exploring, wei2024exploring} contributes to gender disparities within these communities~\cite{vasilescu2014gender}.

\subsection{Study Design}
Our interview protocol was periodically revised as the study evolved, reflecting the iterative nature of qualitative research and the need to explore the meaning of participants' contexts in depth~\cite{creswell2017research, creswell2016qualitative, tan2024more}. This iterative process was essential to ensure that the overall protocol and the specific questions remained aligned with our research goals and effectively surfaced participants' needs and challenges. 

First, participants were asked to share their current model development workflow, and where and why they usually deploy their creations. This provided a foundational understanding of how they build and distribute their AI models, as well as what they prioritize in their workflows. 
Next, participants shared their experiences and opinions regarding how they practice RAI approaches in their workflows, based on the existing approaches proposed by model marketplaces and RAI researchers. We focused on two approaches: (1) a safety guideline for responsible model development~\cite{safetyCivit} and (2) a model attribution tool combined with an AI-generated content (AIGC) indicator. 

\begin{itemize}
    \item \textbf{Responsible Model Development Guideline: } It describes what model creators should consider during the training and testing process of their models. It introduces prohibited image output types (e.g., CSAM, NCII, publicity violation) and prohibited custom data used for their creations (e.g., Do not create resources with the purpose of depicting extreme fetishes or illegal conduct, such as rape, sexual abuse, scat or bestiality.) 
    We used an existing guideline introduced by current model marketplaces as `Responsible Resource Development'~\cite{safetyCivit}. 
     \item \textbf{Model Attribution tool \& AIGC indicator:} Since static model cards can be complicated for non-AI practitioners to engage with~\cite{mitchell2019model, turri2024transparency, tang2025navigating}, we used an improved design inspired by He et al.~\cite{he2025contributions}, originally intended to disclose AI attribution of AIGC.
     In this design, users can select options to disclose upstream dependencies, such as the pre-trained models and datasets they used, and the attribution of their models' creative concepts.
     We also presented an AIGC indicator illustrating what downstream users might see when images are generated using participants' models. Following design conventions from the Coalition for Content Provenance and Authenticity (C2PA)~\cite{contenta}, the indicator displays the original pre-trained model, the creator's creative concept, the creator and model names, and a link to access the model.
\end{itemize}

We presented each RAI approach and asked which, if any, they currently use in their workflows and how they apply them; for those who did not currently adopt these approaches, we walked through how the approaches work. Then, we asked in detail how they would adopt them and what they would expect to achieve through the approaches.

Additionally, beyond evaluating RAI approaches presented, we asked participants to envision a platform regime that prioritizes their personal values. This helped us understand emerging regulatory needs and the potential benefits such a regime would offer their workflows. Interviews were conducted online and an online visual board was used to present probes~\cite{boards}. The detailed content is available at \url{https://osf.io/e6a3p}.

\subsection{Data Analysis}
Interviews were transcribed via Otter.AI~\cite{otter}; Korean interviews (P2, P19) were translated using DeepL~\cite{deepl}. We employed reflexive thematic analysis~\cite{braun2006using, braun2019reflecting} to identify patterns of meaning. In an initial code-alignment workshop, three authors collaboratively sorted and interpreted quotes from three transcripts to develop preliminary themes, guided by the research questions. Based on these initial themes, two authors independently coded the remaining transcripts. When new themes emerged, three authors engaged in discussion to refine them. Finally, the first author reviewed the full thematic structure and reconciled codes to ensure consistency in level of abstraction.

\subsubsection{Ethical Considerations}
To create a comfortable environment where participants could share their perspectives candidly and to minimize social desirability bias, all interviews were conducted online by the first author alone. Before beginning each interview, the researcher explained the purpose of the study and assured participants that there were no appropriate answers, so as not to steer responses toward any predetermined direction. Transcripts were subsequently shared with the research team for collaborative analysis. Throughout the process, we took care not to place participants in ethically uncomfortable situations.

\section{Finding 1: Individual Model Creators' Two Coexisting Logics of Accountability}
\label{find:1}

Individual model creators conceptualized their accountability for their derivatives under two distinct logics — (1) commons-based peer production and (2) personal creative property. 
These two accountability approaches coexisted but could lead to different accountability practices, including how creators identified the downstream impact and harm of their derivative models and how they addressed these concerns in ways that might diverge from platform policies. 

\subsection{Commons-Based Accountability: Reciprocity as the Basis of Responsibility}

Echoing peer production of open-source software and remix cultures~\cite{monroy2011computers, bartsch2024origin}, derivative models are perceived as \textbf{artifacts socially produced through the participation of communities}. P14 framed his work as part of a collective innovation effort with his peers: \textit{``It (building a model) is, at the moment, a small step, whatever we're doing. We will maybe be the best available in our field.''} (P14) In their communities, composed of small groups of creators apart from model marketplaces, creators' work-in-progress and resources (e.g., ComfyUI workflow templates for fine-tuning models) are voluntarily shared, mostly produced based on members' intellectual curiosity. 
Additionally, model modification requires different technical parameters (e.g., datasets, hyperparameters) for different visual outcomes, making peer feedback essential for creators to refine their workflows. Peers' practical feedback, such as resolving error messages and testing reports with their checkpoint files, helps them in the iterative refinement process. Thus, being responsible is a way to consistently remain within their communities and build their models, which depend on others' voluntary contributions. P6 mentioned: \textit{``It (Crediting models I used) helps (increasing their)  visibility and also builds trust in the community.''}

\begin{figure*}[t]
\centering
\includegraphics[width=0.9\textwidth]{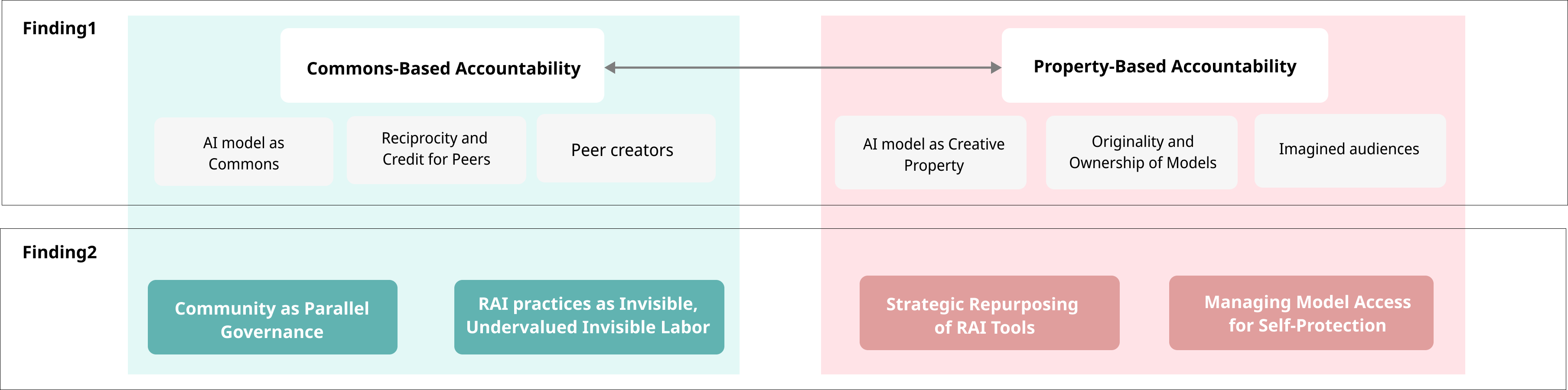} 
\caption{We summarize our findings. Finding 1 examines how creators conceptualize accountability based on their values, while Finding 2 traces how each accountability logic shapes their engagement with self-regulatory Responsible AI approaches.} 
\label{fig1}
\end{figure*}

Considering the impact of their derivative models, creators viewed their accountability as limited to fellow creators within their communities, rather than extending to the broader upstream and downstream stakeholders (e.g., users).
Echoing Widder et al.'s findings~\cite{widder2022limits}, participants considered unexpected use of models by downstream users to be uncontrollable, leaving responsibility to users to behave in good faith. For upstream data owners, whose image data may have been scraped by creators for the fine-tuning process, participants framed their data collection process as legally compliant under \textit{fair use}, by which the scraped images were used as guides rather than resources to be directly taken. P19 mentioned, \textit{``These (derivative) models are transformative in such a way. [...] I think the laws are still developing, and I generally don't feel I am stealing anything, because these models very rarely can reproduce the original work (images) in the first place.''} Within this norm, their approaches in their workflow are limited to follow other community members' data practices, not clearly recognizing the impact of their derivatives in open model ecosystem.

Instead, giving credit when participants distribute their models is a \textit{virtue} that retains motivation for their volunteer work within their community~\cite{benkler2006commons}. `Credit' is attributed to appreciation of other fellow creators' assistance to their models through resources, rather than accurate provenance disclosure when they used others' pipeline templates or checkpoint files (P3, P16). For instance, they encouraged other downstream users to introduce their usernames or unique numbers to signal how their content is attributed. However, these actions are reciprocal implicit expectations across model creators with no rules or guidelines indicating what and how model creators should disclose as attribution (P1, P6, P19). Thus, \textbf{these actions mostly depend on model creators' personal judgment} in calculating how their derivative models are sourced from other model creators' work and datasets shared in communities.

\subsection{Property-Based Accountability: Originality and Ownership as the Basis of Responsibility}

Users of derivative models were often framed as `engineers' contributing to community-based models or company products~\cite{widder2022limits, gorwa2024moderating}. By contrast, in the image-generation model context, \textbf{participants identified themselves as independent ``artists'' or ``creators'' }who build artifacts that generate experimental images using existing ingredients—pre-trained models. P16 said, \textit{``A good model creator is a creative, decisive one.''} Collecting images, pre-trained models to use, and adapters to transfer subjects or styles (e.g., IPAdaptor) requires multiple refinements to gain expected image outputs and these ideas are obtained through multiple individuals' trial and error. Thus, participants considered their derivatives as original artifacts sourced from their ideas. P13 mentioned, \textit{``As time goes by, you have to practice (your) originality, where you can come up with something unique, something original for you.''}

Herein, `originality' was used to express the novelty of their models, compared to other fine-tuned models and pre-trained models they used. The idea for originality began with examining weaknesses of pre-trained models and brainstorming how to modify the models (P2, P5, P10, P16). P10 mentioned, \textit{``when I wanted to generate images of `Hanbok [Korean traditional costume]' (from Stable Diffusion XL), `Hanbok' prompt either produced something entirely unrelated or brought up images of Kimono [Japanese traditional costume]. I collected images of Hanbok and manually tagged each of them with text labels, and then used that dataset to generate images.''} 
As previous researchers noted~\cite{suresh2021framework}, a series of choices and practices driven from their social identities, experiences were embedded in collecting data and choosing concepts. This made them feel as though the models represent their identity. P5, who creates Afro style models, said, \textit{``Most creatives lean towards the Eurocentric side. I wanted something different. I wanted to portray myself in the model.''}

This perception created strong psychological ownership of derivative models, leading them to treat their derivatives as \textbf{potential future income sources}. They wanted to gain full accountability for their derivatives throughout their workflow. Currently, opportunities to earn revenue on model marketplaces are limited~\cite{duan2024they}, but creators believe that their models would eventually function as licensed assets, for which they would be compensated once legal frameworks and technical infrastructure more clearly recognize their technical efforts. P13 mentioned, \textit{``In the long run, it should be mine, and you can't be proud of something if you use [...] someone else's property. That is why I always try to ensure I don't use anyone's licensed property.''}Likewise, their practices were self-protective, rather than an ethical obligation to upstream creators. Similarly, P15 mentioned \textit{``I own the right to use every image that I collected and avoid any copyright issues that may later affect my model when it becomes officially public.''}
They engaged in data practices they believed to be `ethical' in their workflow, which were not validated and were mostly gained from community knowledge-sharing. These include collecting images on their own (e.g., taking pictures by visiting a hair salon — P1), blurring facial identities through AI, generating images using AI image generators (e.g., Leonardo), or choosing images that have no person or creative commons (P15). Given this perception, they described other creators who replicated their creative concepts without attribution as `thieves' and `pirates' who took their benefits (P1, P11, P17, P18).

\textbf{Downstream users who might use their models for image generation or re-use them for another derivative were a central consideration} when participants decided their creative concepts, echoing creative workers who aspire to high visibility on other social media platforms (e.g., YouTubers, Instagrammers) (P10, P14, P17)~\cite{simpson2023rethinking, tafesse2023content}. Their norms are centered on imagined audiences' satisfaction~\cite{duffy2017platform}, negotiating between normative ethics and the users' preferences. P9 mentioned, \textit{``when things are ethical, they (derivative models) are boring.''} He attributed the continued creation of sexually offensive image-generation models to audience demand, noting that audiences often want to generate sexually or politically inciting images. Hence, their focus sometimes ignores IP rights holders' values to maximize audiences' attention. P4 created models that enabled people to generate memes of a specific soccer player: \textit{``Everyone was waiting for him [the Soccer player] to not do it [score], and I created a meme (using my custom model) in case he failed to do it [score]. It really went viral.'' }

\section{Finding 2: Divergent Engagements with Platform Responsible AI Practices}
\label{find:2}

Having identified the two logics through which creators conceptualize accountability for their derivative models, we now turn to how they engage with Responsible AI approaches — safety guidelines and attribution tools — provided by platforms.

\subsection{Creator Community as Parallel Governance to Displace Platform Authority}

Rather than model marketplaces' official policy, \textbf{participants heavily relied on guidance and resources of communities for developing their models,} where model creators are committed to exchanging ideas, sharing their outputs, and collectively gaining feedback outside of those public platforms. P14 mentioned, \textit{``I have a small community on Discord, mostly for testing. I have people there, so when I make a small change, they just test it and tell me, `Okay, now there's a problem here.'\,''} As found in prior literature~\cite{sadek2025challenges}, the static guideline for responsible model development does not provide actionable ways to work for their unique and diverse workflows. P6 mentioned, \textit{``I often feel constrained by vague or overly strict content policies... the platform should clearly define what's allowed and what's not.''} 
Since they often create models that generate experimental images, they needed to devise how to fit the rules to their practices. P15 mentioned, 
\textit{``Learning from other models is really important for me, since it makes me understand and get better prompts for my own model. I can have something to twist to fit in on my model.''}
Thus, from the communities, participants obtained more approachable tips and tricks that they could directly adopt to their models (P1, P3, P13, P16, P17). For instance, P1 shared his Google Colab link in his Discord community channel, then other community members tested them and advised him on how to revise the code.

Such reliance on community is also grounded in an overall distrust of model marketplace platforms. P2 mentioned, \textit{``If more users come in and there's more traffic, the platform succeeds. So I'm somewhat skeptical that platforms can actually solve ethical problems... at first there was no censorship at all (in CivitAI). It didn't matter whether minors came in or not.''} The safety guideline was believed to be a performative and symbolic action that displaces accountability onto individual users. 
Even if they follow the guideline, risks may persist if downstream users are willing to tweak to generate explicit images; since the solution is imperfect, they viewed this guideline as political~\cite{gillespie2010politics}, prioritizing platforms' interests over user support rather than enabling open sharing. 
Participants thought consistent and comprehensive communication with users regarding platforms' policy changes is required to enable creators to adopt the changes quickly. P15 mentioned, \textit{``You see the policies normally change. (When) you wake up, there's this policy that has been introduced by the government that requires everybody.''} From that sense, they believed that their creator community is more reliable than platforms (P10, P11, P14, P17). P17 said \textit{``I always say that it's good we keep the communities as small as possible. It promotes a sense of responsibility.'' }

However, their reliance on community commitment might reflect a skewed viewpoint of community-prioritized norms. Participants followed `the same as what others do', focusing on improving `quality' of image outputs when developing their derivatives (P3, P6, P13, P18). In this sense, they lacked an inclusive perspective in representing cultures, gender, and other dimensions of diversity, leaving some participants unable to find adequate resources within their communities (P2, P16). For instance, most derivative models shared in communities almost exclusively represent women as sexual symbols, making it difficult for P2 to apply community resources to her workflow. This indicates that model creators' responsible development workflows can be established or restructured by what the majority of model creators in their associated community endorse. 

\subsection{RAI Practices as Invisible, Undervalued Labor in Communities}

Although participants recognized the need to reduce potential risks of their models in development process, they perceived practicing RAI approaches as those requiring massive invisible labor that goes unrecognized by platforms and audiences. P3 noted, \textit{``it would be a lot of work on my side to work on such a big thing. But if I successfully execute such a thing, it would be an added advantage to whoever is going to interact with my model.''} 
In open-source software development, code contributors often perform activities (e.g., version control) to operate their projects as active communities, which are often invisible and emotionally intensive~\cite{geiger2021labor}.
Similarly, the effects of RAI practices are rarely visible in their model outputs, meaning that safer models depend on creators' willingness to perform labor whose results others cannot easily see. 
P10 described, \textit{``I don't really compare prompts one by one. Because if I did that too, I would truly have no time left and nothing would get produced.''}

Examining upstream models through audits fell solely on individual creators when those pre-trained models did not provide enough information for participants to judge whether they contain bias (P2, P6). \textit{``When using multiple downloaded checkpoints, it's hard to confirm if the underlying models introduce biases or hidden content that violate platform policies.''} (P6). 
However, their practices were often limited to testing models with prompts that downstream users would likely use (P10).
Participants mentioned that \textit{``Risks are always sleeping''} (P16), questioning whether risks could be fully removed even through rigorous regulation of their development process. Because these practices also required substantial time and money for GPU usage, they considered robust evaluation as \textit{``not ideal.''}
As a result, this labor is perceived as undervalued and overlooked, and practices to mitigate risks were often minimized, left to individual creators to carry out according to their own criteria.

\subsection{Strategic Repurposing of RAI Tools for Creator Visibility}

Given the characteristics of derivative models, participants felt that the attribution tool would not create clear attribution of their upstream models. P2 noted, \textit{``Practically all existing models are scraped from somewhere, so there are no authentic models.''} Because derivative models emerge through complex iterations — involving pipeline templates, custom data scraping, and other creators' resources — participants found it difficult to determine the appropriate scope of attribution for their models. P3 said, \textit{``That person (upstream creators) also brought in someone else's model anyway, so it feels like the open-source spirit. But if I just directly used the thing they're selling as-is, then I think I should credit that person.''} 
Therefore, there are mixed opinions across participants in deciding who and what should be credited. P18 mentioned, \textit{``I think it's also important to include the original data owners. [...] I've seen AI [tools] do this, so other creators that the content has been borrowed from — if it could be acknowledged, because it would have respected the original content owner.''} 

This uncertainty led creators to reinterpret attribution tools and AIGC indicators not as transparency mechanisms, but as marketing features that could enhance the visibility of their derivative models (P6, P9). \textit{``This (tool -- AIGC indicator) would help in discoverability and trust, because when many users mention my page and people generate images using my model, it will increase my [model] discoverability''} (P6). They viewed the indicator as a branding signature that could authenticate their AI-generated images and signal credibility to audiences (P9). 
Thus, although accurate and clear data provenance is important, participants wished to re-organize content to disclose based on how downstream users would be attentive to their models. They were cautious about whether AIGC indicators and their attribution disclosure would negatively impact their model visibility (P2, P3, P7, P17, P18): \textit{``It is important that every piece of information you put as an attribution has a target [audience] and needs to be there. When you put the attribution, it is a good idea to make sure that any information you think is not necessary should not be included.''} (P17) These reactions suggest that creators selectively appropriate attribution tools — disclosing information strategically to maximize visibility rather than transparently conveying model provenance as intended by platform designers.

\subsection{Managing Model Access for Self-Protection}
Participants were concerned that their models might be exploited or targeted by harassment from unidentified downstream users. Open model marketplaces have the advantage of reaching larger audiences, as P12 noted, \textit{``it's way easier for me to access more people there who have the same vision of creating models.''} However, they worried that audiences might criticize them if their models produced negative results. P14 described, \textit{``you can't control everyone's movement, and then it can be traced back to you. I thought the creator is a different entity from the creation, but now it is just coming to me that it can be used against me.''} Once models were uploaded, unexpected downstream use or negative reactions became largely uncontrollable. To mitigate these risks, participants established their own gatekeeping strategies, utilizing platform features such as access controls and disclaimers to delineate accountability boundaries.
P1 described, \textit{``Other model creators cannot replicate my model because I put guidelines — my licensing, watermarks on my work, and also disable downloading my content. It's for read-only.''} 
On the other hand, P10 deliberately obscured model information to reduce discoverability. He scraped social media images of `Hanbok' (traditional Korean attire) for his models, although he recognized this practice verged on infringement: \textit{``Even though the model I made isn't one that could really get me into trouble, I still worried that if I uploaded it there, people might later question how I built my dataset. It wasn't a concrete fear, but more like a lingering concern [...] I just make sure the name isn't associated with `Hanbok' so it doesn't show up in searches. I avoid providing as much information as possible.''}

Five participants shared their models exclusively within private communities through Google Colab or Drive links. These arrangements were grounded in trust built through sustained social interactions — with close school friends, colleagues in similar fields (e.g., fashion, anime fandom), and online acquaintances. P11 described, \textit{``They [people who I share a model with] are my friend[s]. We have been through a lot together. So, yeah, I really trust them.''} Through these relationships, participants could anticipate how members would behave and how they should respond if problems arose from model use. P9 noted, \textit{``I can get direct messages from these people, and understand different individuals according to their characteristics.''}

%
\section{Discussion}

Open model users---people who download, modify, and distribute open models---have received growing attention from researchers seeking to understand how harms are produced in the open model ecosystem~\cite{widder2022limits, tang2025navigating, lee2026open, choksi2025brief}. We focused particularly on individual creators creating image generation models. Our findings showed that contributing to open model development---especially creating derivatives of upstream models---can be motivated not only by enjoyment-based collaboration but also by the personal assetization of one's creations. These varied motivations can lead to different responsible AI practices that are considered acceptable only within particular community norms. In Lee et al.'s work, open model users adopt open models for entrepreneurial benefit, valuing less restrictive licenses and reduced guardrails~\cite{lee2026open}. However, if creators release their modified models as open models without clear data disclosures or attribution practices, such models are no longer truly ``open'' from downstream users' perspectives, since auditing them requires substantial resources. 

\subsection{Individual Creators as a Locus of Risk in the Open Model Ecosystem}

Our findings show that individuals without formal organizational processes, resources, or goals exhibit distinctive behaviors in how they develop derivative models and how their perceived accountability shifts depending on their motivations and creative goals. However, these characteristics may in turn undermine the transparency of their models, introducing risks for downstream dependencies.

First, \textbf{individuals' workflows are largely self-directed}, and continuously adjusted based on their creative visions and peer feedback. In organizational contexts, AI practitioners' workflows are structured, collaborative, and divided~\cite{widder2023dislocated}. They discuss the legitimacy of their practices with internal legal teams or AI compliance experts~\cite{bogucka2024co, balayn2025unpacking}. By contrast, individual creators manage nearly all aspects of their workflow independently, drawing on techniques and resources from community knowledge and cross-platform channels. Their decisions about model development are largely grounded in self-judgment and operate within legal ambiguity (e.g., fair use). 
Thus, unless creators proactively seek ethical guidance for their workflow from communities, unintended dependencies may emerge that obscure critical characteristics of their models, introducing vulnerabilities for downstream users~\cite{hopkins2025ai}.

Moreover, \textbf{individuals' derivative models reflect their personal experiences, value orientations, and domain knowledge.} The choices creators make in collecting, curating, and annotating data, fine-tuning pre-trained models to achieve desired image outputs, and evaluating model performance are all shaped by their individual subjectivity. These situated standpoints can enable derivative models to address socio-cultural needs that general-purpose models often fail to adequately capture — such as fine-tuning for low-resource cultural image representation~\cite{zhang2024low}. However, at the same time, this subjectivity can reproduce dominant community norms when creators lack reflexivity about their own positionality, risking the unintentional embedding of biases through data collection practices and decisions about target image outputs~\cite{winner2017artifacts, scheuerman2021datasets}. As our findings showed (Section~\ref{find:2}), community norms can reproduce skewed representation, reflecting the demographic composition of creator communities. Such biases can be further compounded by upstream pre-trained models that may disproportionately misrepresent or underrepresent particular cultures, identities, and social groups~\cite{shelby2024creative, katzman2023taxonomizing, bird2023typology}.

Additionally, \textbf{individuals' motivations for creating and distributing derivative models differ from those of organizations that adopt open models within specific domains.} Some individuals position themselves as solo entrepreneurs who generate financial value through their creations, embracing the monetization affordances of model marketplace platforms~\cite{bhargava2022creator}. Others view their work as part of remix and `participatory culture'~\cite{jenkins2006fans}, in which creators produce amateur media artifacts and contribute to peer communities. 
This orientation leads them to treat their derivatives as communal objects~\cite{benkler2006commons}, while the broader social consequences of their work often go unexamined, as they frame their activities as recreational and harmless within their peer communities~\cite{jenkins2007confronting}.

Finally, \textbf{individuals' perceived accountability is a moving target;} it is continually negotiated in relation to internal and external factors, including the degree of psychological ownership they hold over their creations~\cite{polimetla2026paradigm}, the imagined audiences whose appreciation they anticipate~\cite{duffy2017platform}, and how they interpret prevailing norms through platform governance~\cite{jakesch2022different}. When actors using AI are situated outside the technology companies that govern its deployment, uncertainty around AI governance and insufficient information about upstream models hinder their ability to enact RAI practices~\cite{kawakami2024responsible}. This challenge emerges for individual creators; the untraceability of downstream use, combined with the high accessibility of model marketplace platforms, generates considerable uncertainty, leading them to prioritize self-protection in their engagement with RAI practices.

Taken together, our findings indicate that individual creators may constitute a significant bottleneck for responsible AI in the open model ecosystem. The risks of open models can compound as models are re-used and re-derived by individuals who are themselves shaped by platforms, communities, and end-users.

\subsection{Policy Implications}

Open model marketplaces face a contested question of regulatory intensity~\cite{gorwa2024moderating}, but given the risks that individual creators may introduce to the open model ecosystem, we argue that open model marketplaces should adopt a more proactive moderation stance, moving beyond their position as neutral intermediaries — particularly for derivative models. Encouragement for practicing RAI approaches alone is insufficient to ensure that individuals exercise them adequately.


\subsubsection{Rewarding Responsible Practice, Not Just Prohibiting Harm}

Not engaging with responsible AI practices is not technically a case of ‘legitimately wronging’ someone. Rather, it relies on individuals’ intrinsic motivation~\cite{sadek2025challenges}. Thus, it should be treated differently from technology-enabled sexual abuse (synthetic NCII, CSAM), which is strongly prohibited on current open model marketplace platforms. Platforms should orient their moderation strategies toward incentivizing creators who contribute to safer model ecosystems by demonstrating their RAI practices, complementing guidelines that focus solely on prohibitions. Positive incentives, such as visibility boosts and attribution recognition, can provide strong motivation for creators seeking broader audience engagement to invest meaningfully in RAI practices~\cite{gillespie2022not, torhonen2020fame}. Concretely, rather than surfacing only quantitative indicators of platform activity (e.g., download counts, follower numbers) on creator profiles, platforms could foreground qualitative indicators of responsible practice — such as creators' documented evaluation approaches and dataset curation methods — as visible signals of accountability work~\cite{marlow2013activity}. However, platforms should also consider the emergence of creators who game the moderation systems to maximize these incentives, or struggle to interpret moderation criteria when they remain opaque~\cite{ma2021advertiser, choi2023creator}. To mitigate these risks, platforms should clearly communicate moderation criteria and mechanisms to creators through transparent, externally-facing policies. 

\subsubsection{Removing Barriers to Creators' RAI Engagement}

Platforms should commit not only to protecting public safety by regulating models that produce harmful outputs, but also to safeguarding creators against the perceived risks that discourage meaningful engagement with RAI practices enhancing model transparency — such as reputational harms from anonymous audiences, unauthorized use of their models, and plagiarism by other creators. 
Enhanced model transparency can expose creators to personal attacks, particularly historically marginalized creators who have long been disproportionately subjected to audience harassment, ranging from identity-based attacks to coordinated negative reviews~\cite{thomas2022s}. Such harassment renders creators vulnerable, degrades their platform experiences, and may ultimately drive them off the platforms entirely. Open model marketplaces, as an emerging context for cultural production, may give rise to distinct forms of these harms, as creators have limited capacity to trace how their derivative models are used or misused after deployment~\cite{widder2022limits}.

\subsubsection{Cultivating Stewardship Culture Among Creators for Responsible Practice}

To prevent unintended harms from individual creators' misjudgment, platforms should cultivate a culture of stewardship in which creators take responsibility for guiding one another toward RAI practices. Stewardship has well-established precedent in open source software culture, where it has been shown to lower maintenance costs, increase flexibility, and improve operational efficiency~\cite{tridgell2025stewardship, priest1997privatization, o2007governance}. While OSS stewardship has traditionally focused on codebase maintenance, the underlying governance structures — distributed authority, peer accountability, and community-defined norms — could be productively adapted to encompass ethical guidance and harm prevention in the open model marketplace context. However, as our findings (Finding 5.1) indicate, community-based governance can also reproduce dominant norms and exclude marginalized perspectives. Stewardship structures therefore require accompanying support for inclusive participation and oversight, such as cross-community auditing, in which creators from different communities periodically review one another's practices to surface blind spots that may go unnoticed within a single community.

\subsubsection{Recognizing Derivative Models in Attribution Framework}

Beyond governance focused on internal stakeholders, platforms should also engage external stakeholders in the open model ecosystem. Existing frameworks for standardizing AI attribution and content authenticity~\cite{contenta, he2025contributions} have largely overlooked individual model creators as relevant stakeholders, as well as the distinctive characteristics of their models. At present, AI companies log digital fingerprints and edit histories as metadata when end-users generate images with AI tools, while social media platforms display AIGC indicators (e.g., ``AI info'') to signal AI-generated content. As these frameworks extend to derivative model contexts, the granularity of modification methods and fine-tuning sources will likely become important dimensions of attribution. When downstream users rely on derivative models to generate images, who should be credited as the primary contributor to the resulting content? This question becomes especially complicated when creators substantially customize pre-trained models — by altering stylistic character or fine-tuning with training data from unverified sources~\cite{hacker2026regulation}. Moreover, if derivative models are documented within AIGC attribution systems, AIGC indicators would shift in function — no longer serving solely as disclosure mechanisms for AI use, but also operating as forms of credit recognizing creators' contributions. 

In light of these complexities, future attribution framework for derivative models should engage model creators directly in defining what modifications warrant disclosure, how their contributions should be represented, and what ethical concerns emerge from their own perspectives.

\subsection{Limitations}

We acknowledge several limitations. First, we focused exclusively on the viewpoints of creators of image generation models, particularly Text-to-Image models. The needs and challenges they encounter to responsibly develop models may vary for different modalities (e.g., text and speech).
Whereas images convey creative ideas visually, making originality easier to recognize, creators of language models may feel a less pressing need for originality.
We also acknowledge that small samples may not fully represent individual creators' voices. Also, our research participants were based in the U.S., so findings may not be globally generalizable.
Owing to differing jurisdictions, users in other countries may have differing perspectives on our results.

\subsection{Conclusion}

Our study surfaces how creators engage with current governance approaches and remedies to support them for Responsible AI, based on their perceptions and experiences.
We believe our findings offer insights into how individuals may act as hidden actors who affect the trust and safety of the model marketplace, and broadly the open model ecosystem, by developing and sharing derivative models on platforms.
We also shed light on what factors shape how these individuals construct their perceived accountability within this emerging yet still niche platform ecosystem.

\section*{Positionality Statement}

Our research team consists of scholars with backgrounds in cultural production, human-AI interaction, and privacy. We have prior experience conducting qualitative research on how AI governance shapes the behavior of social media users, freelance creators, and creative workers. This combined expertise directs our analytical attention to both individual creators' situated experiences and the broader platform structures that shape them. We therefore approached this project with a focus on creators' lived experiences on emerging platforms and on how platform governance operates in relation to those experiences.

We acknowledge that our positioning---as researchers based in U.S. institutions---may shape what we attend to and what we overlook, and that creators in other geographic or cultural contexts may navigate platform governance differently. Throughout the analysis, we worked reflexively to surface tensions between individual creators' practices and the structural conditions imposed by platforms, while remaining alert to how our own disciplinary lenses framed what counted as a meaningful finding.

\section*{Acknowledgements}
We thank our participants for generously sharing their thoughts and experiences with us. We also thank the anonymous reviewers, whose feedback helped us strengthen this work. The first author deeply appreciates Dr. danah boyd, Dr. Karen Levy, Dr. Ned Cooper, and Daniel Mwesigwa in providing their advices to improve this work. We utilized ChatGPT to correct grammar and create Tables in the manuscript. We did not use any AI-driven content.

\bibliography{aaai2026}

@inproceedings{turri2024transparency,
  title={Transparency in the Wild: Navigating Transparency in a Deployed AI System to Broaden Need-Finding Approaches},
  author={Turri, Violet and Morrison, Katelyn and Robinson, Katherine-Marie and Abidi, Collin and Perer, Adam and Forlizzi, Jodi and Dzombak, Rachel},
  booktitle={Proceedings of the 2024 ACM Conference on Fairness, Accountability, and Transparency},
  pages={1494--1514},
  year={2024}
}

@inproceedings{widder2022limits,
  title={Limits and possibilities for “ethical ai” in open source: A study of deepfakes},
  author={Widder, David Gray and Nafus, Dawn and Dabbish, Laura and Herbsleb, James},
  booktitle={Proceedings of the 2022 ACM Conference on Fairness, Accountability, and Transparency},
  pages={2035--2046},
  year={2022}
}

@article{bovens2010two,
  title={Two Concepts of Accountability: Accountability as a Virtue and as a Mechanism},
  author={Bovens, Mark},
  journal={West European Politics},
  volume={33},
  number={5},
  pages={946--967},
  year={2010},
  publisher={Taylor \& Francis}
}

@inproceedings{lee2026open,
  title={Open AI in the Wild: Adoption and Adaptation of Open Models on r/LocalLLaMA},
  author={Lee, Woohyeuk and Howison, James and Lee, Min Kyung and Li, Hanlin},
  booktitle={The 2026 ACM Conference on Fairness, Accountability, and Transparency},
  pages={7203--7236},
  year={2026}
}

@inproceedings{meneely2009secure,
  title={Secure open source collaboration: an empirical study of linus' law},
  author={Meneely, Andrew and Williams, Laurie},
  booktitle={Proceedings of the 16th ACM conference on Computer and communications security},
  pages={453--462},
  year={2009}
}

@article{widder2023open,
  title={Open (for business): Big tech, concentrated power, and the political economy of open AI},
  author={Widder, David Gray and West, Sarah and Whittaker, Meredith},
  journal={Concentrated Power, and the Political Economy of Open AI (August 17, 2023)},
  year={2023}
}

@article{eiras2024position,
  title   = {Position: Near- to Mid-term Risks and Opportunities of Open-source Generative AI},
  author  = {Eiras, Francisco and Petrov, Aleksandar and Vidgen, Bertie and Schroeder de Witt, C. and Pizzati, Fabio and Elkins, Katherine and Mukhopadhyay, Supratik and Bibi, Adel and Botos, Csaba and Steibel, Fabro and others},
  journal = {Journal of Machine Learning Research},
  year    = {2024}
}

@inproceedings{jakesch2022different,
  title={How different groups prioritize ethical values for responsible AI},
  author={Jakesch, Maurice and Bu{\c{c}}inca, Zana and Amershi, Saleema and Olteanu, Alexandra},
  booktitle={proceedings of the 2022 ACM conference on fairness, accountability, and transparency},
  pages={310--323},
  year={2022}
}

@inproceedings{bird2023typology,
  title={Typology of risks of generative text-to-image models},
  author={Bird, Charlotte and Ungless, Eddie and Kasirzadeh, Atoosa},
  booktitle={Proceedings of the 2023 AAAI/ACM Conference on AI, Ethics, and Society},
  pages={396--410},
  year={2023}
}

@incollection{winner2017artifacts,
  title={Do artifacts have politics?},
  author={Winner, Langdon},
  booktitle={Computer ethics},
  pages={177--192},
  year={2017},
  publisher={Routledge}
}

@inproceedings{choi2023creator,
  title={Creator-friendly algorithms: Behaviors, challenges, and design opportunities in algorithmic platforms},
  author={Choi, Yoonseo and Kang, Eun Jeong and Lee, Min Kyung and Kim, Juho},
  booktitle={Proceedings of the 2023 CHI Conference on Human Factors in Computing Systems},
  pages={1--22},
  year={2023}
}

@misc{industry,
  author = {{Meta}},
  title = {Industry {{Leading}}, {{Open-Source AI}} | {{Llama}}},
  howpublished = {https://www.llama.com/},
  year = {2023},
  urldate      = {2025-09-01}
}

@book{creswell2016qualitative,
  title={Qualitative inquiry and research design: Choosing among five approaches},
  author={Creswell, John W and Poth, Cheryl N},
  year={2016},
  publisher={Sage publications}
}

@inproceedings{duffy2017platform,
  title={Platform-specific self-branding: Imagined affordances of the social media ecology},
  author={Duffy, Brooke Erin and Pruchniewska, Urszula and Scolere, Leah},
  booktitle={Proceedings of the 8th international conference on social media \& society},
  pages={1--9},
  year={2017}
}

@article{gillespie2010politics,
  title={The politics of ‘platforms’},
  author={Gillespie, Tarleton},
  journal={New media \& society},
  volume={12},
  number={3},
  pages={347--364},
  year={2010},
  publisher={Sage Publications Sage UK: London, England}
}

@article{torhonen2020fame,
  title={Fame and fortune, or just fun? A study on why people create content on video platforms},
  author={T{\"o}rh{\"o}nen, Maria and Sj{\"o}blom, Max and Hassan, Lobna and Hamari, Juho},
  journal={Internet Research},
  volume={30},
  number={1},
  pages={165--190},
  year={2020},
  publisher={Emerald Publishing Limited}
}

@article{sadek2025challenges,
  title={Challenges of responsible AI in practice: scoping review and recommended actions},
  author={Sadek, Malak and Kallina, Emma and Bohn{\'e}, Thomas and Mougenot, C{\'e}line and Calvo, Rafael A and Cave, Stephen},
  journal={AI \& society},
  volume={40},
  number={1},
  pages={199--215},
  year={2025},
  publisher={Springer}
}

@misc{ykilcher,
  author       = {Kilcher, Yannic},
  title        = {ykilcher/gpt-4chan},
  year         = {2022},
  howpublished = {\url{https://huggingface.co/ykilcher/gpt-4chan}},
  note         = {Accessed: 2025-09-03},
  urldate      = {2025-09-01}
}

@article{gorwa2024moderating,
  title={Moderating model marketplaces: Platform governance puzzles for AI intermediaries},
  author={Gorwa, Robert and Veale, Michael},
  journal={Law, Innovation and Technology},
  volume={16},
  number={2},
  pages={341--391},
  year={2024},
  publisher={Taylor \& Francis}
}

@article{laufer2025anatomy,
  title={Anatomy of a Machine Learning Ecosystem: 2 Million Models on Hugging Face},
  author={Laufer, Benjamin and Oderinwale, Hamidah and Kleinberg, Jon},
  journal={arXiv preprint arXiv:2508.06811},
  year={2025}
}

@article{parker2019snowball,
  title={Snowball sampling},
  author={Parker, Charlie and Scott, Sam and Geddes, Alistair},
  journal={SAGE research methods foundations},
  year={2019},
  publisher={Sage}
}

@misc{maiberg2025payment,
  author       = {Maiberg, Emanuel},
  title        = {Payment Processors Are Pushing {AI} Porn Off Its Biggest Platforms},
  year         = {2025},
  howpublished = {\url{https://www.404media.co/payment-processors-are-pushing-ai-porn-off-its-biggest-platforms/}},
  urldate      = {2025-07-14},
  note         = {404 Media}
}

@misc{otter,
  author       = {{Otter.ai}},
  title        = {Otter Voice Meeting Notes},
  year         = {2025},
  howpublished = {\url{https://otter.ai/home}},
  urldate      = {2025-09-01}
}

@article{tang2025navigating,
  title={Navigating Uncertainties: Understanding How GenAI Developers Document Their Models on Open-Source Platforms},
  author={Tang, Ningjing and Li, Megan and Winecoff, Amy and Madaio, Michael and Heidari, Hoda and Shen, Hong},
  journal={arXiv preprint arXiv:2503.23574},
  year={2025}
}

@misc{safetyCivit,
  author       = {{Civit AI}},
  title        = {Safety Center: Policies and Guidelines},
  year         = {2025},
  howpublished = {\url{https://civitai.com/safety}},
  langid       = {english}
}

@inproceedings{lima2025public,
  title={Public Opinions About Copyright for AI-Generated Art: The Role of Egocentricity, Competition, and Experience},
  author={Lima, Gabriel and Grgi{\'c}-Hla{\v{c}}a, Nina and Redmiles, Elissa M},
  booktitle={Proceedings of the 2025 CHI Conference on Human Factors in Computing Systems},
  pages={1--32},
  year={2025}
}

@article{marchal2024generative,
  title={Generative AI misuse: A taxonomy of tactics and insights from real-world data},
  author={Marchal, Nahema and Xu, Rachel and Elasmar, Rasmi and Gabriel, Iason and Goldberg, Beth and Isaac, William},
  journal={arXiv preprint arXiv:2406.13843},
  year={2024}
}

@inproceedings{mitchell2019model,
  title={Model cards for model reporting},
  author={Mitchell, Margaret and Wu, Simone and Zaldivar, Andrew and Barnes, Parker and Vasserman, Lucy and Hutchinson, Ben and Spitzer, Elena and Raji, Inioluwa Deborah and Gebru, Timnit},
  booktitle={Proceedings of the conference on fairness, accountability, and transparency},
  pages={220--229},
  year={2019}
}

@inproceedings{he2025contributions,
  title={Which contributions deserve credit? perceptions of attribution in human-ai co-creation},
  author={He, Jessica and Houde, Stephanie and Weisz, Justin D},
  booktitle={Proceedings of the 2025 CHI Conference on Human Factors in Computing Systems},
  pages={1--18},
  year={2025}
}

@inproceedings{simpson2023rethinking,
  title={Rethinking creative labor: A sociotechnical examination of creativity \& creative work on TikTok},
  author={Simpson, Ellen and Semaan, Bryan},
  booktitle={Proceedings of the 2023 CHI Conference on Human Factors in Computing Systems},
  pages={1--16},
  year={2023}
}

@misc{boards,
  author       = {{Miro}},
  title        = {Boards in Education Team - Miro},
  year         = {2025},
  howpublished = {\url{https://miro.com/app/dashboard/}},
  urldate      = {2025-09-01}
}

@article{braun2019reflecting,
  title={Reflecting on reflexive thematic analysis},
  author={Braun, Virginia and Clarke, Victoria},
  journal={Qualitative research in sport, exercise and health},
  volume={11},
  number={4},
  pages={589--597},
  year={2019},
  publisher={Taylor \& Francis}
}

@article{braun2006using,
  title={Using thematic analysis in psychology},
  author={Braun, Virginia and Clarke, Victoria},
  journal={Qualitative research in psychology},
  volume={3},
  number={2},
  pages={77--101},
  year={2006},
  publisher={Taylor \& Francis}
}

@article{vasilescu2014gender,
  title={Gender, representation and online participation: A quantitative study},
  author={Vasilescu, Bogdan and Capiluppi, Andrea and Serebrenik, Alexander},
  journal={Interacting with Computers},
  volume={26},
  number={5},
  pages={488--511},
  year={2014},
  publisher={Oxford University Press}
}

@article{tan2025if,
  title={If open source is to win, it must go public},
  author={Tan, Joshua and Vincent, Nicholas and Elkins, Katherine and Sahlgren, Magnus},
  journal={arXiv preprint arXiv:2507.09296},
  year={2025}
}

@inproceedings{liesenfeld2024rethinking,
  title={Rethinking open source generative AI: open-washing and the EU AI Act},
  author={Liesenfeld, Andreas and Dingemanse, Mark},
  booktitle={Proceedings of the 2024 ACM Conference on Fairness, Accountability, and Transparency},
  pages={1774--1787},
  year={2024}
}

@inproceedings{langenkamp2022open,
  title={How open source machine learning software shapes ai},
  author={Langenkamp, Max and Yue, Daniel N},
  booktitle={Proceedings of the 2022 AAAI/ACM Conference on AI, Ethics, and Society},
  pages={385--395},
  year={2022}
}

@inproceedings{paris2025opening,
  title={Opening the Scope of Openness in AI},
  author={Paris, Tamara and Moon, AJung and Guo, Jin LC},
  booktitle={Proceedings of the 2025 ACM Conference on Fairness, Accountability, and Transparency},
  pages={1293--1311},
  year={2025}
}

@article{kilamo2012proprietary,
  title={From proprietary to open source—Growing an open source ecosystem},
  author={Kilamo, Terhi and Hammouda, Imed and Mikkonen, Tommi and Aaltonen, Timo},
  journal={Journal of Systems and Software},
  volume={85},
  number={7},
  pages={1467--1478},
  year={2012},
  publisher={Elsevier}
}

@article{widder2024open,
  title={Why ‘open’AI systems are actually closed, and why this matters},
  author={Widder, David Gray and Whittaker, Meredith and West, Sarah Myers},
  journal={Nature},
  volume={635},
  number={8040},
  pages={827--833},
  year={2024},
  publisher={Nature Publishing Group UK London}
}

@inproceedings{marlow2013activity,
  title={Activity traces and signals in software developer recruitment and hiring},
  author={Marlow, Jennifer and Dabbish, Laura},
  booktitle={Proceedings of the 2013 conference on Computer supported cooperative work},
  pages={145--156},
  year={2013}
}

@inproceedings{monroy2011computers,
  title={Computers can't give credit: How automatic attribution falls short in an online remixing community},
  author={Monroy-Hern{\'a}ndez, Andr{\'e}s and Hill, Benjamin Mako and Gonzalez-Rivero, Jazmin and Boyd, Danah},
  booktitle={Proceedings of the SIGCHI Conference on Human Factors in Computing Systems},
  pages={3421--3430},
  year={2011}
}

@inproceedings{kawakami2024responsible,
  title={Do responsible AI artifacts advance stakeholder goals? four key barriers perceived by legal and civil stakeholders},
  author={Kawakami, Anna and Wilkinson, Daricia and Chouldechova, Alexandra},
  booktitle={Proceedings of the AAAI/ACM Conference on AI, Ethics, and Society},
  volume={7},
  pages={670--682},
  year={2024}
}

@inproceedings{berman2024scoping,
  title={A Scoping Study of Evaluation Practices for Responsible AI Tools: Steps Towards Effectiveness Evaluations},
  author={Berman, Glen and Goyal, Nitesh and Madaio, Michael},
  booktitle={Proceedings of the 2024 CHI Conference on Human Factors in Computing Systems},
  pages={1--24},
  year={2024}
}

@misc{contenta,
  author       = {{Content Authenticity Initiative}},
  title        = {Content Credentials},
  year         = {2024},
  howpublished = {\url{https://www.contentcredentials.org/}},
  note         = {Accessed: 2025-09-03},
  urldate      = {2025-09-05}
}

@article{gebru2021datasheets,
  title={Datasheets for datasets},
  author={Gebru, Timnit and Morgenstern, Jamie and Vecchione, Briana and Vaughan, Jennifer Wortman and Wallach, Hanna and Iii, Hal Daum{\'e} and Crawford, Kate},
  journal={Communications of the ACM},
  volume={64},
  number={12},
  pages={86--92},
  year={2021},
  publisher={ACM New York, NY, USA}
}

@misc{deepl,
  author       = {{DeepL SE}},
  title        = {DeepL Translate: The World's Most Accurate Translator},
  year         = {2025},
  howpublished = {\url{https://www.deepl.com/en/translator}},
  urldate      = {2025-09-01}
}

@inproceedings{bogucka2024co,
  title={Co-designing an AI impact assessment report template with AI practitioners and AI compliance experts},
  author={Bogucka, Edyta and Constantinides, Marios and {\v{S}}{\'c}epanovi{\'c}, Sanja and Quercia, Daniele},
  booktitle={Proceedings of the AAAI/ACM Conference on AI, Ethics, and Society},
  volume={7},
  pages={168--180},
  year={2024}
}

@inproceedings{balayn2025unpacking,
  title={Unpacking Trust Dynamics in the LLM Supply Chain: An Empirical Exploration to Foster Trustworthy LLM Production \& Use},
  author={Balayn, Agathe and Yurrita, Mireia and Rancourt, Fanny and Casati, Fabio and Gadiraju, Ujwal},
  booktitle={Proceedings of the 2025 CHI Conference on Human Factors in Computing Systems},
  pages={1--20},
  year={2025}
}

@inproceedings{choksi2025brief,
  title={The Brief and Wondrous Life of Open Models},
  author={Choksi, Madiha Zahrah and Mandel, Ilan and Benthall, Sebastian},
  booktitle={Proceedings of the 2025 ACM Conference on Fairness, Accountability, and Transparency},
  pages={3224--3240},
  year={2025}
}

@inproceedings{pepe2024hugging,
  title={How do hugging face models document datasets, bias, and licenses? an empirical study},
  author={Pepe, Federica and Nardone, Vittoria and Mastropaolo, Antonio and Bavota, Gabriele and Canfora, Gerardo and Di Penta, Massimiliano},
  booktitle={Proceedings of the 32nd IEEE/ACM International Conference on Program Comprehension},
  pages={370--381},
  year={2024}
}

@article{shelby2024creative,
  title={Creative ML assemblages: the interactive politics of people, processes, and products},
  author={Shelby, Renee and Srinivasan, Ramya and Burgdorf, Katharina and Lena, Jennifer C and Rostamzadeh, Negar},
  journal={Proceedings of the ACM on Human-Computer Interaction},
  volume={8},
  number={CSCW1},
  pages={1--30},
  year={2024},
  publisher={ACM New York, NY, USA}
}

@book{jenkins2006fans,
  title={Fans, bloggers, and gamers: Exploring participatory culture},
  author={Jenkins, Henry},
  year={2006},
  publisher={nyu Press}
}

@article{jenkins2007confronting,
  title={Confronting the challenges of participatory culture--media education for the 21st century (Part Two)},
  author={Jenkins, Henry},
  journal={Nordic Journal of Digital Literacy},
  volume={2},
  number={2},
  pages={97--113},
  year={2007}
}

@article{scheuerman2021datasets,
  title={Do datasets have politics? Disciplinary values in computer vision dataset development},
  author={Scheuerman, Morgan Klaus and Hanna, Alex and Denton, Remi},
  journal={Proceedings of the ACM on human-computer interaction},
  volume={5},
  number={CSCW2},
  pages={1--37},
  year={2021},
  publisher={ACM New York, NY, USA}
}

@article{bhargava2022creator,
  title={The creator economy: Managing ecosystem supply, revenue sharing, and platform design},
  author={Bhargava, Hemant K},
  journal={Management Science},
  volume={68},
  number={7},
  pages={5233--5251},
  year={2022},
  publisher={INFORMS}
}

@inproceedings{hopkins2025ai,
  title={Ai supply chains: An emerging ecosystem of ai actors, products, and services},
  author={Hopkins, Aspen and Cen, Sarah H and Struckman, Isabella and Ilyas, Andrew and Videgaray, Luis and M{\k{a}}dry, Aleksander},
  booktitle={Proceedings of the AAAI/ACM Conference on AI, Ethics, and Society},
  volume={8},
  number={2},
  pages={1266--1277},
  year={2025}
}

@article{tafesse2023content,
  title={Content creators' participation in the creator economy: Examining the effect of creators’ content sharing frequency on user engagement behavior on digital platforms},
  author={Tafesse, Wondwesen and Dayan, Mumin},
  journal={Journal of Retailing and Consumer Services},
  volume={73},
  pages={103357},
  year={2023},
  publisher={Elsevier}
}

@inproceedings{katzman2023taxonomizing,
  title={Taxonomizing and measuring representational harms: A look at image tagging},
  author={Katzman, Jared and Wang, Angelina and Scheuerman, Morgan and Blodgett, Su Lin and Laird, Kristen and Wallach, Hanna and Barocas, Solon},
  booktitle={Proceedings of the AAAI Conference on artificial intelligence},
  volume={37},
  number={12},
  pages={14277--14285},
  year={2023}
}

@book{creswell2017research,
  title={Research design: Qualitative, quantitative, and mixed methods approaches},
  author={Creswell, John W and Creswell, J David},
  year={2017},
  publisher={Sage publications}
}

@article{schneider2025investigating,
  title={Investigating toxicity and Bias in stable diffusion text-to-image models},
  author={Schneider, Matthias and Hagendorff, Thilo},
  journal={Scientific Reports},
  volume={15},
  number={1},
  pages={31401},
  year={2025},
  publisher={Nature Publishing Group UK London}
}

@misc{maiberg_huggingface_2025,
  author       = {Maiberg, Emanuel},
  title        = {Hugging Face Is Hosting 5{,}000 Nonconsensual {AI} Models of Real People},
  howpublished = {\emph{404 Media}},
  year         = {2025},
  month        = jul,
  day          = {15},
  url          = {https://www.404media.co/},
  note         = {Accessed: 2026-05-18}
}

@inproceedings{wei2024exploring,
  title={Exploring the use of abusive generative AI models on Civitai},
  author={Wei, Yiluo and Zhu, Yiming and Hui, Pan and Tyson, Gareth},
  booktitle={Proceedings of the 32nd ACM International Conference on Multimedia},
  pages={6949--6958},
  year={2024}
}

@article{widder2023dislocated,
  title={Dislocated accountabilities in the “AI supply chain”: Modularity and developers’ notions of responsibility},
  author={Widder, David Gray and Nafus, Dawn},
  journal={Big Data \& Society},
  volume={10},
  number={1},
  pages={20539517231177620},
  year={2023},
  publisher={SAGE Publications Sage UK: London, England}
}

@inproceedings{song2024exploring,
  title={Exploring the Potential of Novel Image-to-Text Generators as Prompt Engineers for CivitAI Models},
  author={Song, Sophia and Song, Joy and Lee, Junha and Kang, Younah and Moon, Hoyeon},
  booktitle={2024 16th IIAI International Congress on Advanced Applied Informatics (IIAI-AAI)},
  pages={626--631},
  year={2024},
  organization={IEEE}
}

@article{raffel2023building,
  title={Building machine learning models like open source software},
  author={Raffel, Colin},
  journal={Communications of the ACM},
  volume={66},
  number={2},
  pages={38--40},
  year={2023},
  publisher={ACM New York, NY, USA}
}

@inproceedings{tan2024more,
  title={More than model documentation: uncovering teachers' bespoke information needs for informed classroom integration of ChatGPT},
  author={Tan, Mei and Subramonyam, Hari},
  booktitle={Proceedings of the 2024 CHI Conference on Human Factors in Computing Systems},
  pages={1--19},
  year={2024}
}

@inproceedings{polimetla2026paradigm,
  title={A Paradigm for Creative Ownership},
  author={Polimetla, Tejaswi and Gero, Katy Ilonka and Glassman, Elena L},
  booktitle={Proceedings of the 2026 CHI Conference on Human Factors in Computing Systems},
  pages={1--16},
  year={2026}
}

@article{hacker2026regulation,
  title={The regulation of fine-tuning: Federated compliance for modified general-purpose AI models},
  author={Hacker, Philipp and Holweg, Matthias},
  journal={Computer Law \& Security Review},
  volume={60},
  pages={106234},
  year={2026},
  publisher={Elsevier}
}

@inproceedings{zhang2024low,
  title={Low-resource vision challenges for foundation models},
  author={Zhang, Yunhua and Doughty, Hazel and Snoek, Cees GM},
  booktitle={Proceedings of the IEEE/CVF Conference on Computer Vision and Pattern Recognition},
  pages={21956--21966},
  year={2024}
}

@inproceedings{suresh2021framework,
  title={A framework for understanding sources of harm throughout the machine learning life cycle},
  author={Suresh, Harini and Guttag, John},
  booktitle={Proceedings of the 1st ACM Conference on Equity and Access in Algorithms, Mechanisms, and Optimization},
  pages={1--9},
  year={2021}
}

@article{geiger2021labor,
  title={The labor of maintaining and scaling free and open-source software projects},
  author={Geiger, R Stuart and Howard, Dorothy and Irani, Lilly},
  journal={Proceedings of the ACM on human-computer interaction},
  volume={5},
  number={CSCW1},
  pages={1--28},
  year={2021},
  publisher={ACM New York, NY, USA}
}

@article{anisuzzaman2025fine,
  title={Fine-tuning large language models for specialized use cases},
  author={Anisuzzaman, DM and Malins, Jeffrey G and Friedman, Paul A and Attia, Zachi I},
  journal={Mayo Clinic Proceedings: Digital Health},
  volume={3},
  number={1},
  pages={100184},
  year={2025},
  publisher={Elsevier}
}

@book{birch2020assetization,
  title={Assetization: Turning things into assets in technoscientific capitalism},
  author={Birch, Kean and Muniesa, Fabian},
  year={2020},
  publisher={MIT press}
}

@article{benkler2006commons,
  title={Commons-based peer production and virtue},
  author={Benkler, Yochai and Nissenbaum, Helen},
  journal={Journal of political philosophy},
  volume={14},
  number={4},
  year={2006}
}

@misc{darka,
  author       = {{WIRED}},
  title        = {The Dark Side of Open Source AI Image Generators},
  year         = {2025},
  howpublished = {\url{https://www.wired.com/story/dark-side-open-source-ai-image-generators/}},
  urldate      = {2025-12-15}
}

@misc{googlea,
  author       = {{Google Public Policy Blog}},
  title        = {The Meaning of Open},
  year         = {2009},
  howpublished = {\url{https://publicpolicy.googleblog.com/2009/12/meaning-of-open.html}},
  urldate      = {2025-11-16}
}

@inproceedings{sola2026,
author = {Canet Sola, Mar and Guljajeva, Varvara},
title = {Why Open Small AI Models Matter for Interactive Art},
year = {2026},
isbn = {9798400720017},
publisher = {Association for Computing Machinery},
address = {New York, NY, USA},
url = {https://doi.org/10.1145/3773699.3789655},
doi = {10.1145/3773699.3789655},
booktitle = {Proceedings of the 2025 12th International Conference on Digital and Interactive Arts: Media Art Cultures, Communities \& Territories},
articleno = {26},
numpages = {8},
location = {
},
series = {ARTECH '25}
}

@article{o2007governance,
  title={The governance of open source initiatives: what does it mean to be community managed?},
  author={O’Mahony, Siobh{\'a}n},
  journal={Journal of Management \& Governance},
  volume={11},
  number={2},
  pages={139--150},
  year={2007},
  publisher={Springer}
}

@article{priest1997privatization,
  title={The privatization of regulation: Five models of self-regulation},
  author={Priest, Margot},
  journal={Ottawa L. Rev.},
  volume={29},
  pages={233},
  year={1997},
  publisher={HeinOnline}
}

@inproceedings{tridgell2025stewardship,
  title={‘Stewardship’as a Fair, Accountable and Transparent Model for Free and Open-Source Software Governance? Looking Beyond the EU’s Cyber Resilience Act},
  author={Tridgell, Jennifer and Singh, Jatinder},
  booktitle={Proceedings of the 2025 ACM Conference on Fairness, Accountability, and Transparency},
  pages={473--484},
  year={2025}
}

@article{ma2021advertiser,
  title={" How advertiser-friendly is my video?": YouTuber's Socioeconomic Interactions with Algorithmic Content Moderation},
  author={Ma, Renkai and Kou, Yubo},
  journal={Proceedings of the ACM on Human-Computer Interaction},
  volume={5},
  number={CSCW2},
  pages={1--25},
  year={2021},
  publisher={ACM New York, NY, USA}
}

@inproceedings{thomas2022s,
  title={“It’s common and a part of being a content creator”: Understanding How Creators Experience and Cope with Hate and Harassment Online},
  author={Thomas, Kurt and Kelley, Patrick Gage and Consolvo, Sunny and Samermit, Patrawat and Bursztein, Elie},
  booktitle={Proceedings of the 2022 CHI conference on human factors in computing systems},
  pages={1--15},
  year={2022}
}

@misc{maiberg2025huggingface,
  author       = {Maiberg, Emanuel},
  title        = {Hugging {F}ace {I}s {H}osting 5,000 {N}onconsensual {AI} {M}odels of {R}eal {P}eople},
  howpublished = {\url{https://www.404media.co/hugging-face-is-hosting-5-000-nonconsensual-ai-models-of-real-people/}},
  year         = {2025},
  month        = jul,
  note         = {404 Media. Accessed: 2026-05-21}
}

@article{gillespie2022not,
  title={Do not recommend? Reduction as a form of content moderation},
  author={Gillespie, Tarleton},
  journal={Social Media+ Society},
  volume={8},
  number={3},
  pages={20563051221117552},
  year={2022},
  publisher={SAGE Publications Sage UK: London, England}
}

@article{harris2023regulate,
  title={How to regulate unsecured “Open-Source” AI: No exemptions},
  author={Harris, David Evan},
  journal={Tech Policy Press},
  year={2023}
}

@article{franccois2025different,
  title={A Different Approach to AI Safety: Proceedings from the Columbia Convening on Openness in Artificial Intelligence and AI Safety},
  author={Fran{\c{c}}ois, Camille and P{\'e}ran, Ludovic and Bdeir, Ayah and Dziri, Nouha and Hawkins, Will and Jernite, Yacine and Kapoor, Sayash and Shen, Juliet and Khlaaf, Heidy and Klyman, Kevin and others},
  journal={arXiv preprint arXiv:2506.22183},
  year={2025}
}

@inproceedings{bartsch2024origin,
  title={The origin and opportunities of developers’ perceived code accountability in open source ai software development},
  author={Bartsch, Sebastian Clemens and Lother, Moritz and Schmidt, Jan-Hendrik and Adam, Martin and Benlian, Alexander},
  booktitle={Proceedings of the AAAI/ACM Conference on AI, Ethics, and Society},
  volume={7},
  number={1},
  pages={94--106},
  year={2024}
}

@article{duan2024they,
  title={" They've Stolen My GPL-Licensed Model!": Toward Standardized and Transparent Model Licensing},
  author={Duan, Moming and Zhao, Rui and Jiang, Linshan and Shadbolt, Nigel and He, Bingsheng},
  journal={arXiv preprint arXiv:2412.11483},
  year={2024}
}

@inproceedings{duan2025position,
  title={Position: Current model licensing practices are dragging us into a quagmire of legal noncompliance},
  author={Duan, Moming and Du, Mingzhe and Zhao, Rui and Wang, Mengying and Wu, Yinghui and Shadbolt, Nigel and He, Bingsheng},
  booktitle={Forty-second International Conference on Machine Learning Position Paper Track},
  year={2025}
}

@article{aksulu2010comprehensive,
  title={A comprehensive review and synthesis of open source research},
  author={Aksulu, Altay and Wade, Michael R},
  journal={Journal of the Association for Information Systems},
  volume={11},
  number={11},
  pages={6},
  year={2010}
}

\end{document}